\documentstyle[12pt]{article}
\begin{document}
\let\a=\alpha \let\b=\beta  \let\c=\chi
\let\d=\delta  \let\e=\varepsilon
\let\f=\varphi \let\g=\gamma \let\h=\eta
\let\k=\kappa  \let\l=\lambda
\let\m=\mu   \let\n=\nu   \let\o=\omega
\let\p=\pi  \let\ph=\varphi
\let\r=\rho  \let\s=\sigma \let\t=\tau
\let\th=\vartheta
\def\hf{{\hat\f}}
\def\hf1{{\hat\f_1}}
\let\y=\upsilon \let\x=\xi \let\z=\zeta
\def\dpr{\partial}
\let\F=\Phi
\let\D=\Delta \let\F=\Phi  \let\G=\Gamma
\let\L=\Lambda \let\Th=\Theta
\let\O=\Omega \let\P=\Pi   \let\Ps=\Psi
\let\Si=\Sigma \let\X=\Xi
\let\Y=\Upsilon
\def\bP{{\bf P}}
\def\TT{{\cal T}}
\def\FF{{\cal F}}
\def\PP{{\cal P}}
\let\io=\infty
\def\pp_F{{\vec p}_F}
\def\V#1{\vec#1}
\def\oo{{\vec \o}} \def\OO{{\vec \O}} \def\uu{{\vec \y}}
\def\xx{{\vec x}} 
\def\F{{\Phi}} 
\def\xxx{\underline{\vec x}} 
\def\yy{{\vec y}} \def\kk{{\vec k}}
\def\pp{{\vec p}}
\def\BB{{\cal B}}
\def\nn{\nonumber}
\def\VV{{\cal V}}
\def\E{{\cal E}} \def\ET{{\cal E}^T}
\def\LL{{\cal L}}\def\RR{{\cal R}}\def\SS{{\cal S}}
\def\NN{{\cal N}}
\def\HH{{\cal H}}\def\DD{{\cal D}}\def\GG{{\cal G}}
\def\ie{{\it i.e.}}

%
%
%
\def\ins#1#2#3{\vbox to0pt{\kern-#2 \hbox{\kern#1 #3}\vss}\nointerlineskip}
%
%
%
\newdimen\xshift \newdimen\xwidth \newdimen\yshift
 
\def\insertplot#1#2#3#4#5{\par%
\xwidth=#1 \xshift=\hsize \advance\xshift by-\xwidth \divide\xshift by 2%
\yshift=#2 \divide\yshift by 2%
\line{\hskip\xshift \vbox to #2{\vfil%
#3 \includegraphics{#4.pst}}\hfill \raise\yshift\hbox{#5}}}
 
\def\initfig#1{%
\catcode`\%=12\catcode`\{=12\catcode`\}=12
\catcode`\<=1\catcode`\>=2
\openout13=#1.ps}
 
\def\endfig{%
\closeout13
\catcode`\%=14\catcode`\{=1
\catcode`\}=2\catcode`\<=12\catcode`\>=12}

\openin14=\jobname.aux \ifeof14 \relax \else
\input \jobname.aux \closein14 \fi
\openout15=\jobname.aux
\title{Peierls instability with electron-electron interaction: the
commensurate case.}
\author{{Vieri Mastropietro}}
\maketitle
\begin{abstract}
We consider a quantum many-body model describing
a system of electrons interacting with themselves
and hopping from one ion to another of a one dimensional
lattice. We show that the ground state energy of such system,
as a functional of the ionic configurations, has local
minima in correspondence of configurations
described by smooth ${\pi\over p_F}$ periodic functions,
if the interaction is repulsive and large enough
and $p_F$ is the Fermi momentum of the electrons.
This means physically that
a $d=1$ metal develop a periodic distortion of its
reticular structure (Peierls instability). 
The minima are found solving the Eulero-Lagrange equations
of the energy by a contraction 
method.
\end{abstract}
\section{Main results}
\subsection{Introduction}
In 1955 Peierls [P] suggested that in a 
one dimensional metal it would be energetically favorable
to develop a periodic distortion of the linear lattice
with period ${p_F\over\pi}$ where $p_F$ is the
{\it Fermi momentum} of the conduction electrons; 
indeed periodic lattice distortions
with period ${p_F\over\pi}$ have been observed experimentally
in many anisotropic compounds [F].
Theoretical attempts
to understand Peierls instability are based
on a {\it variational approach}; a functional
of the ion configurations 
describing the total energy of the metal
is introduced, and if such functional
has a minimum 
in correspondence of ${\pi\over p_F}$-periodic 
configuration, one says that there
is Peierls instability. In [KL] 
the case $p_F={\pi\over 2}$ was considered
and it was indeed shown
that in the {\it Holstein model}, describing
fermions interacting with the ions of the lattice
{\it without} electron-electron interaction, 
the energy has a global minimum in correspondence of a
configuration
of period ${\pi\over p_F}=2$.
The case $p_F={\pi\over 2}$ is rather special
as the hamiltonian enjoys many special symmetries. In [BGM] it
was treated the Holstein model 
with $p_F=\pi{P\over Q}$, 
with $P,Q$ relative prime integers,
and it was shown that the Holstein model energy has local
minima which correspond to ${\pi\over p_F}$ periodic function, if
the interaction is $\le 
e|\log Q|^{-1}$ for some small $\e$.
The Holstein model with large interaction 
was treated in [AAR],
where Peierls instability was proved (in such regime however the
minimizing periodic function has infinite many discontinuities,
contrary to the small interaction case in which it is a smooth function).

There are two main open problems; the first
is what happens considering ${p_F\over\pi}$
as an {\it irrational} number (the {\it incommensurate} case), and the second
is what is the effect of
an interaction among electrons on Peierls
instability.
In this paper we consider the second problem;
its relevance is quite clear as any
realistic model must include some interaction
between electrons. The only paper considering
Peierls instability in presence of an electron-electron
interaction is
[LN], which is limited to the case
$p_F={\pi\over 2}$ so that one can use symmetries
which are absent in the general case.
We will consider the general (spinless) 
{\it commensurate} case $p_F=\pi{P\over
Q}$ with 
{\it weak}
electron-ion and electron-electron interaction. 
No assumptions is done on the relative size as
both the cases are interesting.
There are physical
situations in which the electron-ion interaction
dominates over the electron-electron interaction (like in conventional
low temperature superconductors) 
but other in which the opposite situation
is found (like in high-temperature superconducting cuprates,
see [FJ]). 

\subsection{Peierls instability}

We can assume than that the units of the crystal form
a liner chains with spacing $a=1$ and the fermions,
in the tight binding approximation, are hopping
from one site to another. If $\L=1,2,....,L$ and
$\psi_x^\pm$ are creation
or annihilation spinless fermionic operators 
with periodic boundary conditions
verifying canonical anticommutation relations
the hamiltonian is,
\begin{eqnarray}\label{1}
&&H_F=\sum_{x\in\L}
[t_{x}\psi^+_{x}\psi^-_{x+1}+t_x\psi^+_{x+1}\psi^-_{x}
-(\mu+\nu)\psi^+_x\psi^-_x]+\nn\\
&&+U \sum_{x \in\L}
[\psi^+_{x}\psi^-_x-{1\over 2}][\psi^+_{x+1}\psi^-_{x+1}-{1\over 2}]
\end{eqnarray}
where $\mu$ is
the chemical
potential and $\mu=1-\cos p_F$. 
We assume moreover that $L=i Q$
for $i>0$ integer (this is for preserving
periodic boundary conditions, see below).
The hopping matrix element $t_x$ is a resonance integral between
orbitals at site $x$ and $x+1$ and therefore it depends on the distance
$\phi_x$ between these sites; we can choose
\begin{equation}
t_x=-1-\l\phi_x
\end{equation}
Although in more refined models the lattice distortion
should be treated quantum-mechanically a more common choice,
which we will follow here, is to treat them in the Born-Oppenhmeier
approximation by adding to the fermionic hamiltonian a term 
of the form $\sum_x {\f_x^2\over 2}$.
The total energy of the system is then
\begin{equation}
F(\f)={1\over2}\sum_{x\in\L}\f^2_x+E_0(\f)\label{vf}
\end{equation}
where $E_0(\f)$ is the ground state energy 
of the fermionic hamiltonian $H_F$ \ie
\begin{equation}
E_0(\f)=\lim_{\b\to\io} {Tr H_F e^{-\b H_F}\over Tr e^{-\b H_F}}
\end{equation}
We fix units so that (assuming $p_F\not=0,\pi$)
$\sin p_F=1$.
The last term in the hamiltonian takes into account
the interaction between
fermions with coupling $U$; one site interaction
is forbidden by Pauli principle (the fermions are spinless)
so a nearest neighbor interaction is considered. Of
course $U>0$ in order to describe Coulomb interaction,
but also $U<0$ is conceivable as a consequence of the interaction
with phonons. Finally $\nu$ is a {\it counterterm}
to be fixed to a suitable way as a function of $U$ (see below).

We will consider $F(\f)$ as a functional $F:\O\to R$ 
where $\O$ is the class of functions defined below.

{\bf Definition 1.} {\it $\O$ is the class of functions $\f_x:\L\to R$
periodic with period ${\pi\over p_F}$, even and such that
\begin{equation}
\f_x=\sum_{n=-[Q/2]\atop n\not=0}^{[(Q-1)/2]} \hat\f_n
e^{i 2 p_F n x} \label{2} 
\end{equation}
with $\l\hat\f_1=\s$, $\hat\f_{n}=\hat\f_{-n}$ and
\begin{equation}\label{ccc}
|\l\hat\f_n|\le {C|\s|\over 1+|n|^2}
\end{equation}
with $|\s|<1$. We define a norm in $\O$ as $||\f||=
\sum_{x\in\L}|\f_x|$}

The functional $F(\f)$ is defined over the ${\pi\over p_F}$-periodic
functions. As $x\in \L$ then the Fourier coefficients of $\f_x$  
are $O(Q)$; if $p_F={\pi\over 2}$ there is only one coefficient
$\f_x=\hat\f_1 (-1)^x$. 
By (\ref{ccc})
we are requiring 
that $\f_x$ has the 
$2 p_F$ harmonic non vanishing (by such a condition
if $\hat\f_1=0$ then $\f_x$
is a constant).  
Moreover if $\f$ is an extremal point of $F(\f)$, it must satisfy the
condition $\hat\f_0=\l\r$ where $\r$ is the fermionic density.
On the other hand, 
we can always include $\hat\f_0$ in the chemical potential $\mu$
and then we can
restrict our search of
local minima of $F(\f)$ to fields $\f$ with zero mean.
The infinite length limit is performed along a
sequence of $L$ so that periodic boundary conditions on
$\f_x$ are preserved.  
In the following the $N$ depending constants
(not dependent of $\l, U,Q$)
are generically denoted by $C$; all the other constants
are $N$ independent. With the above notation
we will prove the following theorem.
\vskip.5cm
{\bf Theorem 1}{\it Given the Hamiltonian (\ref{1}) with
$p_F=\pi{P\over Q}$ and $P,Q$ positive prime,
it is possible to find an $\e$ and
a function $\nu=\nu(U)=O(U)$ such that: 

a)If $U\ge 0$ (repulsive interaction) 
and $|\l|,|U|\le {\e\over|\log Q| C}$ 
then $F:\O\to R$ has a local minimum $\f\in\O$
given by (\ref{2})
with
\begin{eqnarray}\label{sp}
&&|\l\hat\f_1|=A(1+g_2)(1+{\h\over\l^2 g_1})^{-{1\over\h}}\nn\\
&&|\l\hat\f_n|\le {C\l^{2N}\over|n|^N}\qquad n\not=1
\end{eqnarray}
where $\h(U)=\b_1 U+O(U^2)$, $g_2=O(\l,U)$
and $g_1(U)=a_1+O(\l,U)$ are continuous functions,
$N$ is a positive integer and $a,\b_1, C$ are positive 
constants.

b)If $U<0$ (attractive interaction) 
but ${|U|\over\l^2 a_1}\le {1\over 2}$
then, if $|\l|,|U|\le {\e\over|\log Q| C}$,
$F:\O\to R$ has a local minimum $\f\in\O$
given by (\ref{2}) (\ref{sp}).

c)If $U<0$ (attractive interaction) 
but ${|U|\over\l^2 a_1}\ge 3$
then, if $|\l|,|U|\le {\e\over|\log Q| C}$, 
there is 
no local minima $\f\in\O$ for $F:\O\to R$}
\vskip.5cm
In the $U=0$ case Peierls instability is found [BGM] with 
$\nu=0$, $|\l|\le
{\e\over |\log Q| C_N}$ and $|\l\hat\f_1|=Ae^{-{a+O(\l)\over\l^2}}$.
The fact that $\nu=0$ means that there is a simple relation between
the chemical potential $\mu$ and the period of the minimizing
$\f_x$, \ie $\mu=1-\cos p_F$. If $U\not=0$ things are different.
To find a minimizing $\f_x$ with period
${\pi\over p_F}$ one fixes the chemical potential
to the value $1-\cos p_F+\nu$, where $\nu$ is a 
suitable counterterm. In some special case 
$\nu=0$ (for instance when $p_F={\pi\over 2}$) but in general
$\nu$ is a non vanishing function.
The above theorem says that, when $U$ is smaller than $\l$ the electron-electron interaction
does not modify essentially the Peierls instability, as we find
that the dependence of $\s$ from $\l$ is essentially
the same as in the $U=0$ case, as
$$|\s|=A[(1+g_2)({1+{\h\over\l^2 g_1}})]^{-{1\over\h}}=
A e^{-{a\over\l^2}[1+O({U\over\l^2})]}$$
On the other hand if $U$ is larger than
$\l$ the electron-electron interaction has a quite dramatic effect.
For attractive interaction
$U<0$ there is {\it no} Peierls instability (\ie the electron-electron
interaction destroys Peierls instability)
while for repulsive interaction $U>0$ there Peierls instability 
but the dependence of $\f_x$ from $\l,U$
is completely different with respect to the $U=0$ case.
Note also that
there is a weak dependence of $\l,U$ on $Q$
\ie $\l,U\le O(\log Q^{-1})$; this means that
our result can be applied to "almost incommensurate"
$\f_x$, \ie such that $Q$ is quite large (but $\log Q$
is reasonably small).
The expansion for the Ground state energy
is similar to the one
discussed in literature in many papers (see for istance [BM])
expect that 1)one has to understand the dependence on $Q$  
of the constants, and it is crucial to have careful bounds
in order to have results valid for  
$\l,U\le O(\log Q^{-1})$; b)one has to prove that $\nu$ is
$\l$-independent. 
Finally note that
the presence of the spin should not change the 
result when $p_F\not={\pi\over 2}$ and $U>0$.

\subsection{The contraction method}

It is well known (see [BM]) that $E_0(\f)$
can be written as a {\it Grassman integral} \footnote{
We use the same symbol $\psi$ for
field and Grassman variables with a traditional abuse of notation.}
\begin{equation}\label{la}
E_0(\f)=-\lim_{\b\to\io}{1\over L\b}\log \int P(d\psi) e^{-UV-\l P-\nu N},
\end{equation}
where
\begin{eqnarray}\label{la1}
&&V=\int_{-{\b\over 2}}^{\b\over 2} d x_{0}
\sum_{x\in\L}
[\psi^+_{\xx}\psi^-_{\xx}-{1\over 2}][\psi^+_{\xx+1}\psi^-_{\xx+1}-{1\over
2}],\\
&&P=-\int_{-{\b\over 2}}^{\b\over 2} d x_{0} \sum_{x\in\L}\f(x)
\psi^+_{\xx}\psi^-_{\xx},\quad N=
\int_{-{\b\over 2}}^{\b\over 2} d x_{0} \sum_{x\in\L}
\psi^+_{\xx}\psi^-_{\xx},\nn
\end{eqnarray}
and $\xx=(x_0,x)$ and $\xx+1=(x_0,x+1)$. $P(d\psi)$ is a
{\it Grassmanian integration} defined on monomials by the anticommutative
Wick rule with propagator
\begin{equation}
g(\xx;\yy)={1\over\b L}\sum_\kk{e^{-i\kk(\xx-\yy)}\over -i k_0-\cos k+\cos
p_F},
\end{equation}
where $\kk=(k_0,k)$.
(\ref{la}) has a well defined $L,\b\to\io$
limit only if the counterterm $\nu$ is chosen
in a suitable way as a function of the parameters
appearing in the Hamiltonian so that the Fermi momentum
is just $p_F$.  

In order to find
the minima of $F(\f)$ we have to differentiate with
respect to $\hat\f_x$, so one has in principle
to take into account the possible dependence
of $\nu$ from $\hat\f_x$, which is in general
very complicated.
However we will show in that it is possible
to choose $\nu$ as {\it independent} from $\l$ and so on $\hat\f_x$. 
This is due to the fact that the chemical potential can
be moved inside the gap opened by $\f_x$ without
affecting any physical property, and we can use
this freedom to fix $\nu$ as independent of $\hat\f_x$.
It follows that a necessary condition for $\f_x\in\O$ 
to be a local minimum for $F(\f)$ is that
it verifies
$\f_x=\l\r_x$ where
$\r_x=\lim_{\b\to\io,\t\to 0}{1\over L}S^{L,\b}(x,\t;x,0)$
and $S^{L,\b}(x,\t;x,0)$ is the Schwinger function defined by,
if $\phi^\pm_\xx$ are Grassman variables
and writing $\int d\xx=\int_{-\b\over 2}^{\b\over 2}\sum_{x\in\L}$,
\begin{equation}
S^{L,\b}(\xx;\yy)={\partial^2\over\phi^+_\xx\partial\phi^-_\yy}
\log\int P(d\psi)e^{-\VV(\psi)-\int
d\xx[\phi^+_\xx\psi_\xx^-+\phi^-_\xx\psi_\xx^+]}|_{\phi=0},
\end{equation}
and $\VV=U V+ \l P+\nu N$. Note that when $L$ is finite $F(\f)$ 
can be considered as an $L$-dimensional function, as $\f_x$
is determined by the values of $\f_1,\f_2,...\f_L$. 
We will see that $F:\O\to R$ is differentiable 
so that
the minima of $F:\O\to R$ verify, 
for $n\not=0$, $n=-[Q/2],...,[(Q-1)/2]$, verify
\begin{equation}\label{6}
\hat\f_n=\l\hat\r_n(\f)
\end{equation}
and $M_{n,m}$ positive definite, where
$$M_{n,m}=\d_{n,m}-\l {\partial\over\partial\hat\f_{m}}\hat\r_n(\f)$$
and $\r(x)=\sum_{n=-[Q/2]\atop n\not=0}^{[(Q-1)/2]} \r(x) 
e^{i 2 p_F n x}$
where
\begin{equation}
\r(x)=\lim_{\b\to\io} {Tr \psi^+_x\psi^-_x e^{-\b H_F}\over Tr e^{-\b H_F}}
\end{equation}
\vskip.5cm
(\ref{6}) are the {\it Eulero-Lagrange} equations
of our variational problem.
\vskip.5cm
We will see in \S 2.3,\S 2.4 that (\ref{6}) can be written as
\begin{equation}\label{bon}
\hat\f_n = - \l^2 c_n(\s) \hat\f_n + \l \tilde \r_n(\s,\F) \; , \quad
\s=\l\hat\f_1\; ,\quad \F=\{\l\hat\f_n\}_{|n|>1}
\end{equation}
where $c_n(\s)$ depends on $\f$ only through
$\s$. 

The equation (\ref{bon}) has of course the trivial solution $\hat\f_n=0,
\forall n$,
but it is easy to see that this is not a local minimum.
Therefore we shall look for solutions such that
$\s\neq 0$, so that we can rewrite (\ref{bon}) as
\begin{eqnarray}\label{4}
&& (1+\l^2 c_1(\s)) = {\l^2\tilde\r_1(\s,\F) \over \s } \nn\\
&& \F_n= \l \hat\f_n = {\l^2 \tilde\r_n(\s,\F) \over (1+\l^2 c_n(\s))} \; ,
\qquad |n|> 1
\end{eqnarray}
Explicit bounds for the quantities appearing in the above equation
are given in the following lemma. 
\vskip.5cm
{\bf Lemma 1} {\it If $\f_x\in\O$
there exists a (Q-independent) $\e$
and a $\h(U),\nu(U)$ independent from $\l$
such that, $|\s|\le {\e^2\over Q^4}$ and $|U|\le\e$
then
\begin{eqnarray}\label{ma}
&&|\tilde\r_n|\le {C |\s|\over |n|^N}\quad |n|\not=0\nn\\
&&|c_n(\s)|\le C |\log Q| Q^{C |U|}\quad |n|\not=1\\
&&c_1(\s)={1\over\h}[
\left({|\s|\over A}\right)^{-\h}-1][a_1+U F]+\s f\nn
\end{eqnarray}
and $|F|,|\tilde\r_1|, |f|\le C$
and $\h=\b U+O(U^2)$}
\vskip.5cm
We find a solution of (\ref{4}) 
considering $\s$ as a variable.
We solve the second of (\ref{4}) by considering $\s$ 
as a parameter and assuming $\l^2,|U|\le {\e\over C_N log Q}$.
Fixed $L=L_i$, $\F$ is a finite sequence of $Q-3$ elements, which can be
thought as a vector in $R^{Q-3}$, which is a function of $\s$.
We consider
the space $\FF=C^1(R^{Q-3})$ of $C^1$-functions of $\s$
with values
in $R^{Q-3}$; the solutions of eq(\ref{4}) can be seen
as fixed points of the operator ${\bf T}_{\l}: \FF\to \FF$,
defined by the equation:
$$[{\bf T}_{\l}(\F)]_n(\s) =
{\l^2 \tilde\r_n(\s,\F(\s)) \over (1+\l^2 c_n(\s))}$$

We shall define, for each positive integer $N$, a norm in $\FF$ in
the following way:
$$||\F||_\FF= sup_{|n|>1} \left\{
|n|^N \left[ |\s|^{-1} |\F_n(\s)|
+\left|{\partial\F_n\over \partial\s}(\s)\right|\right] \right\} 
$$

We shall also define
$$\BB = \{\F\in\FF : \|\F\|_\FF\le 1\}\; $$
$$R(\F)_n(\s) = \tilde\r_n(\s,\F(\s))\;,\quad |n|\ge 2$$

It is an easy corollary of lemma 1 that, if
if $\F,\F'\in\BB$ and
then
$$||R(\F)-R(\F')||_\FF \le C_N
||\F-\F'||_\FF$$
and
$$\|R(0)\|_\FF \le C
\sup_{|n|>1} \left\{ |n|^N |\s|^{|n|\over 10}
\right\}$$

It follows that

{\bf Lemma 2.} {\it There are $\e,c,K$, independent of
$N$, such that for $\l,|U|\le {\e\over  C log Q}$
there exists a unique solution $\F\in\BB$ of the second
of (\ref{4}); moreover the solution satisfies the bound
$$\|\F\|_\FF \le \l^{2N}\; $$
}

{\it Proof of Lemma.}
If $\l^2,|U|\le {\e\over C log Q}$ then from (\ref{ma})
$$\l^2 |c_n(\s)|\le C\l^2|\log Q| Q^{C  |U|}\le 2\e$$
so that for $\e$ small enough
$${\l^2\over 1+\l^2 c_n(\s)}\le 2\l^2$$
so that 
$$\|{\bf T}_{\l}(\F)\|_\FF \le C_N \l^{2}\le 1$$
which means that $\BB$ is invariant under the action
of ${\bf T}_{\l}$.
Moreover if $\F,\F'\in\BB$ for $\l$ small enough
$$\|{\bf T}_{\l}(\F)-{\bf T}_{\l}(\F')\|_\FF \le C
\l^2|\F-\F'|_\FF \le
{1\over 2} |\F-\F'|_\FF$$
so that $T$ is a contraction on $\BB$.
Hence, by the contraction mapping principle, there is
a unique fixed point $\bar\F$ of ${\bf T}_{\l}$ in $\BB$, which can
be obtained as the limit of the sequence $\F^{(k)}$ defined through the
recurrence equation $\F^{(k+1)}={\bf T}_{\l}(\F^{(k)})$, with any initial
condition $\F^{(0)}\in\BB$. If we choose $\F^{(0)}=0$, we get
$$|\bar\F|_\FF \le \sum_{i=1}^\io \|\F^{(i)}-\F^{(i-1)}\|_\FF
\le \sum_{i=1}^\io {1\over 2^{i-1}} \|\F^{(1)}\|_\FF \le
||\F^{(1)}||_\FF$$
On the other hand
$$\|\F^{(1)}\|_\FF = \|{\bf T}_{\l}(0)\|_\FF \le
C_N
\left(\l^2\right)^N$$
which immediately implies that there is a solution belonging
to $\O$.
\vskip.5cm
We insert in the first of (\ref{4})
$\hat\f_n$ as a function of $\s$ and 
we call simply
$$\tilde\r_1(\s,\hat\f_n(\s))\equiv\r_1(\s)$$
and we look for a solution $\s$. 
\vskip.5cm
{\bf Lemma 3} {\it Assume that
$|U|, \l^2 \le {\e\over C \log Q}$.

1)If $U>0$, or if $U<0$ but
${|U|\over\l^2 a_1}<{1\over 2}$
then there exists
a solution of 
$\s=\l^2\r_1(\s)$ such that $|\s| Q^4\le\e^2$

2)If ${|U|\over\l^2 a_1}>{3}$ but $U<0$
there is no solution such that $|\s|\le 1$.}
\vskip.5cm
{\bf Proof of lemma 3}
We write the self-consistence equation
as
$${\h\over\l^2}=[\left({A\over|\s|}\right)^\h-1]
[a^{-1}+U F]+\s
\tilde f$$
where $\h, F$ are $\l$-independent and $|F|,|\tilde f|\le C$. 
We look for a solution of 
the form 
$$|\s|=
A[(1+g_2)({1+{\h\over\l^2 (a_1+U F) }})]^{-{1\over\h}}$$
with $g_2=g_2(\l,U)$ and $O(\e)$. Substituing
by the implicit function theorem there is a solution for
$g_2=O(\e)$ if $({1+{\h\over\l^2 (a^{-1}+U f_1) }})\le {1\over 2}$;
this is true assuming that $U\ge 0$ or $U<0$
and ${a|\h|\over\l^2}\le {1\over 2}$ or ${a|\h|\over\l^2}\ge 3$.

Finally we check that $|\s| Q^4\le\e^2$ if
$|U|, \l^2 \le {\e\over \log Q}$.
Note in fact that if 
$U>0$, $U\le {\e\over\log Q}$
then, for $\e$ small enough 
$$|\s|=A({1+{\h\over\l^2}})^{-{1\over\h}}
= A e^{-{1\over\l^2}[{\l^2\over U}\log(1+{\h\over\l^2})} 
\le A e^{-{1\over\l^2}}\le {1\over Q^{1\over\e}}\le {\e^2\over Q^4}$$
The same bound is true if $U<0$, ${\h\over\l^2}\le {1\over 2}$.

On the other hand if $|{U\over\l^2}|>{3}$,
$U<0$, $U\le {\e\over\log Q}$ surely
$|\s|\ge 2^{{1\over \e}}>2$
for $\e$ small enough.
\vskip.5cm
Theorem 1 is an easy consequence of the above lemmas.

\section{Renormalization Group analysis}
\subsection{The effective potential}
We start 
by studying the {\sl free energy} 
\begin{equation}
E_{L,\b}= - {1\over L\b}
\log \int P(d\psi^{[h,0]})e^{-V(\psi^{[h,0]})}
\end{equation}
with $\lim_{\b\to\io} E_{L,\b}$ giving the ground state energy.
The integration can be performed iteratively by a slight modification
of the procedure described in [BM].
It is convenient to decompose the Grassman
integration $P(d\psi)$ into a product 
of independent integrations.
Let be $|\kk|=\sqrt{k_0^2+||k||_T^2}$.
We write 
\begin{eqnarray}
&&g(\kk)=f_1(\kk) g(\kk)+(1-f_1(\kk)) g(\kk)=\nn\\
&&g^{(u.v.)}(\kk)+g^{(i.r.)}(\kk)\label{3.3}
\end{eqnarray}
where $f_1(\kk)=1-\chi(k-p_F,k_0)-\chi(k+p_F,k_0)$ and
$\chi(k',k)$ is a $C^\io$ function
with compact support such that it is $1$ 
for $|\kk'|\le{a_0\over\g}$ and $0$ for 
$|\kk'|>a_0$, where $\g>1$ and $\g,a_0$ are chosen
so that $\chi(k\pm p_F,k_0)$ are non vanishing only
in two non overlapping regions around $\pm p_F$.
We write $k=k'+\o p_F$, $\o=\pm 1$ and
$$g^{(i.r.)}(\kk)=\sum_{\o=\pm 1}\sum_{h=-\io}^0 f_h(\kk')g(\kk)\equiv
\sum_{\o=\pm 1}\sum_{h=-\io}^0 g^{(h)}(\kk)$$
where $f_h(\kk')=\chi(\g^{-h}\kk')-\chi(\g^{-h+1}\kk')$
has support $O(\g^h)$. Then 
we can see $P(d\psi)$ as the product of independent integrations of fields
$$P(d\psi)=P(d\psi^{(1)})\prod_{h=-\io}^0 \prod_{\o=\pm 1}
P(d\psi^{(h)}_\o)$$
and $\psi^\s_\kk=\psi^{\s(1)}_\kk+\sum_{h=-\io}^0 \sum_{\o=\pm 1}
\psi^{\s(h)}_{\kk,\o}$. The field $\psi^{\s(h)}_{\kk,\o}$
has momenta distant $O(\g^h)$ to $\o p_F$. The idea is to integrate
iteratively $\psi^{(1)},\psi^{(0)},\psi^{(-1)}.....$
going closer and closer to the Fermi surface.
The integration of $\psi^{(1)}$, the {\it ultraviolet} integration,
gives $$e^{-\VV^{(0)}(\psi^{(\le 0)})}=
\int P(d\psi^{(1)}) e^{-\VV^{(0)}(\psi^{(1)}+\psi^{(\le 0)})}$$
where (for shortening the notation ${1\over\b L}\sum_{\kk}$
will be denoted by $\int d\kk$)
\begin{eqnarray}
&&\VV^{(0)}(\psi^{(\le 0)})=
\sum_{n=1}^\io\sum_{m=0}^\io \int d\kk_1...d\kk_{2n}
\psi^{(\le 0)\s_1}_{\kk_1}...
\psi^{(\le 0)\s_n}_{\kk_{2n}}\nn\\
&&W_{2n,m}^{0}(\kk_1,...,\kk_{2n})\d(\sum_{i=1}^{2n}\s_i\kk_i+2m \pp_F)
\label{v}
\end{eqnarray}
where if $\pp_F=(p_F,0)$, $\sigma_i=\pm$ and the kernels
$W_{n,m}^{0}(\kk_1,...,\kk_n;z)$ are $C^\io$ bounded functions such
that $W^0_{n,m}=W^0_{n,-m}$ and $|W^0_{n,m}|\le C^n z^{\max(2,n/2-1)}$ if
$z=Max(|\l|,|U|,|\nu|)$. By an explicit computation
it follows that $W^0_{4,0}=U+O(U^2)$, $W^0_{4,m}=O(U\s)$ for
$m\not=0$ and $W^0_{2,m}=\s+O(\s U)$ for $m\not=0$.
$\VV^{(0)}$ is called {\it effective potential at scale $0$};
note that it contains non local interactions between an arbitrary
number of fermions.

The study of the {\it infrared} integration is much more involved.
The integration is performed iteratively, setting $Z_0=1$, $\s_0=\s$,
in the following way: once that the fields $\psi^{(0)},...,\psi^{(h+1)}$
have
been integrated we have
\begin{equation}
\int P_{Z_h}(d\psi^{(\le h)}) \, 
e^{-\VV^{(h)}(\sqrt{Z_h}\psi^{(\le h)})}\label{3.22}
\end{equation}
with $C_h(\kk')^{-1}=\sum_{j=-\io}^h f_j(\kk')$ and
$\a(k')=(\cos k'-1)\cos p_F$, $v_0=\sin p_F$:

\begin{eqnarray}
&&P_{Z_{h}}(d\psi^{(\le h)}) = 
\prod_{\kk'}\prod_{\o=\pm1} d\psi^{(\le h)+}_{\kk'+\o\pp_F,\o}
d\psi^{(\le h)-}_{\kk'+\o\pp_F,\o}\label{3.10b}\\
&&\exp \Big\{ -\sum_{\o=\pm1} \int d\kk'
C_h(\kk') Z_{h}
\Big[\Big( -ik_0-\a(k') +\o v_0\sin k' \Big)\nn\\
&&\psi^{(\le0)+}_{\kk'+\o\pp_F,\o}
\psi^{(\le0)-}_{\kk'+\o\pp_F,\o}
- \sigma_{h}(\kk') \, \psi^{(\le0)+}_{\kk'+\o\pp_F,\o}
\psi^{(\le0)-}_{\kk'-\o\pp_F,-\o} \Big]
\Big\}\nn
\end{eqnarray}
Note that after $h$ steps the integration is different with
respect to the initial one; there is a wave function renormalization
$Z_h$ and a mass term $\s_h$.
Moreover the effective potential at scale $h$ has the form
\begin{eqnarray}\label{18j}
&&\VV^{(h)}(\psi^{(\le h)})=
\sum_{n=0}^\io\sum_{m=0}^\io
\int d\kk_1...d\kk_{2n}\prod_{i=1}^n \psi^{\sigma_i
(\le h)}_{\kk'_i+\o_i \pp_F,\o_i}\label{ty}\\
&&d(\sum_{i=1}^{2n} \sigma_i(\kk'_i+\o_i
\pp_F)+2m\pp_F)
W_{n,m}^{h}(\kk'_1+\oo_1 \pp_F,..;,..;,..;,..;,..;,..;,..;\{\o\})\nn
\end{eqnarray}
In order to integrate $\psi^{(h)}$ we write
$\VV^{(h)}$ as $\LL \VV^{(h)}+\RR \VV^{(h)}$, with $\RR=1-\LL$.
The $\LL$ operation is defined to extract the non irrelevant terms
in $\VV^{(h)}$; it is easy to check from a power counting argument 
that the terms in $\VV^{(h)}$ involving six
or more fields are irrelevant then $\LL=0$ on such terms.
Moreover we will define $\LL=0$
on the addenda in eq(\ref{ty}) 
{\it not} verifying the condition
\begin{equation}
\sum_{i=1}^{2n}\s_i\o_i p_F+2 m p_F=0 \quad {\rm mod.}\; 2\pi\label{f1}
\end{equation}
which means that we are considering irrelevant the terms
such that the sum of momenta measured from the Fermi surface
is not vanishing (but it can be arbitrary small, due to the
irrationality of ${p_F\over\pi}$). In conclusion
the definition of $\LL$ is the following

1) If $2n>4$ then
$$\LL W_{2n,m}^{h}(\kk_1,...)=0$$

2) If $2n=4$ then
\begin{equation}
\LL W_{4,m}^{h}(\kk_1,...)
=\d_{m,0}\d_{\sum_{i=1}^{4}\s_i\oo_i,0} 
W_{4,m}^{h}(\o_1  \pp_F,...,
\o_4 \pp_F)\label{loc1}
\end{equation}

3)If $2n=2$, $\o_1=\o_2$ then
\begin{eqnarray}
&&\LL\{ 
W_{2,m}^{h}(\kk'_1+\o_1 \pp_F,
\kk'_2+\o_2 \pp_F)=\d_{m ,0}
[W_{2,m}^{h}(\o_1 \pp_F,\o_2 \pp_F)\nn\\
&&+\o_1E(k'+\o_1 p_F)\partial_{k}
W_{2,m}^{h}(\o_1 \pp_F,\o_2 \pp_F)+\nn\\
&&k^0 \partial_{k_0}W_{2,m}^{h}(\o_1 \pp_F,\o_2 \pp_F)]\label{loc2}
\end{eqnarray}
where $E(k'+\o p_F)=v_0\o\sin k'+(1-\cos k')\cos p_F$ and the symbol
$\partial_k,\partial_{k_0}$ means discrete derivatives. 

4)If $n=2$, $\o_1=-\o_2$ then
\begin{equation}\label{loc3}
\LL
W_{2,m}^{h}(\kk'_1+\o_1 \pp_F,
\kk'_2+\o_2 \pp_F)
=\d_{m ,\o_2}W_{2,m}^{h}(\o_1 \pp_F,\o_2 \pp_F)
\end{equation}

The Kronecker deltas in the r.h.s. of (\ref{loc1}),
(\ref{loc2})(\ref{loc3}) ensure that $\LL=0$ if (\ref{f1})
is not verified.
We find
$$
\LL\VV^{(h)}(\psi)=\g^h n_h F_\nu^{(\le h)}+
s_h F_\s^{(\le h)}+z_h F_\z^{(\le h)}+a_h
F_\a^{(\le h)}+u_h F_U^{(\le h)}$$
where
\begin{eqnarray}
&&F_\s^{(\le h)}=\sum_{\o=\pm 1} \int d\kk'
\psi^{(\le h)+}_{\kk'+\o \pp_F,\o}
\psi^{(\le h)-}_{\kk'-\o \pp_F,-\o}\nn\\
&&F_i^{(\le h)}=\sum_{\o=\pm 1}\int d\kk' f_i(\kk')
\psi^{(\le h)+}_{\kk'+\o \pp_F,\o}
\psi^{(\le h)-}_{\kk'+\o \pp_F,\o}\nn\\
&&
F_U^{(\le h)}=\int [\prod_{i=1}^4
d\kk'_i]\d(\sum_{i=1}^4\s_i\kk_i)\times\nn\\
&&\psi^{(\le h)+}_{\kk'_1+\pp_F,1}
\psi^{(\le h)-}_{\kk'_2+\pp_F,1}
\psi^{(\le 0)+}_{\kk'_3-\pp_F,-1}
\psi^{(\le 0)-}_{\kk'_4-\pp_F,-1}\d(\sum_{i=1}^4\s_i\kk_i)\nn
\end{eqnarray}
where $i=\nu,\z,\a$ and $f_\nu=1$, $f_\z=-i k_0$
and $f_\a=E(k'+\o p_F)$; moreover
$u_0=U(\hat
v(0)-\hat v(2 p_F))+O(U^2)$,
$s_0=O(U \l)$, 
$a_0,z_0=O(U^2)$,
$n_0=\nu+O(U)$. 
Note that in $\LL V^{(h)}$ there are terms
renormalizing mass and the wave function renormalization and
it is convenient to include them in the fermionic
integration writing
\begin{equation}
\int P_{Z_h}(d\psi^{(\le h)}) \, e^{-\VV^{(h)}(\sqrt{Z_h}\psi^{(\le
h)})}=
\int \tilde P_{Z_{h-1}}(d\psi^{(\le h)}) \,
e^{-\tilde\VV^{(h)}(\sqrt{Z_h}\psi^{(\le h)})}
\label{3.22a}
\end{equation}
where
$\tilde  P_{Z_{h-1}}(d\psi^{(\le h)})$ is defined as 
$P_{Z_{h}}(d\psi^{(\le h)})$ eq(\ref{3.10b}) with $Z_{h-1}$ and $\s_{h-1}$
replacing $Z_h,\s_h$, with
$$Z_{h-1}(\kk')=Z_h(1+C_h^{-1}(\kk')z_h)\quad Z_{h-1}(\kk')
\s_{h-1}(\kk')=Z_h(
\s_h(\kk')+C_h^{-1}(\kk') s_h$$ 
Moreover 
$$\tilde
\VV^{(h)}=\LL\tilde\VV^{(h)}+(1-\LL)\VV^{(h)}$$ and
\begin{equation}\label{3.11b}
\LL\tilde\VV^{(h)}=\g^h n_h F_\nu^{(\le h)}+(a_h-z_h)
F_\a^{(\le h)}+u_h F_U^{(\le h)}
\end{equation}

The r.h.s of (\ref{3.22a}) can be written as
\begin{equation}\label{ml1}
\int P_{Z_{h-1}}(d\psi^{(\le h-1)}) \int \tilde
P_{Z_{h-1}}(d\psi^{(h)}) \, e^{-\tilde \VV^{(h)}(\sqrt{Z_h}\psi^{(\le
h)})}
\end{equation}
where $ P_{Z_{h-1}}(d\psi^{(\le h-1)})$ and $\tilde
P_{Z_{h-1}}(d\psi^{(h)})$ are given
by eq(\ref{3.10b}) with $Z_{h-1}(\kk')$ replaced by $Z_{h-1}(0)\equiv
Z_{h-1}$
and
$C_h(\kk')$ replaced with
$C_{h-1}(\kk')$
and $\tilde f_h^{-1}(\kk')$ respectively, if
$$\tilde f_h(\kk')=Z_{h-1}[{C_h^{-1}(\kk')\over Z_{h-1}(\kk')}-
{C_{h-1}^{-1}(\kk')\over Z_{h-1}}]$$
and $\psi^{(\le h)}$ replaced with
$\psi^{(\le h-1)}$ and $\psi^{(h)}$ respectively. Note that $\tilde
f_h(\kk')$
is a compact support function, with support of width $O(\g^h)$
and far $O(\g^h)$ from the "singularity" \ie $\o p_F$.
The Grassmanian integration $\tilde P_{Z_{h-1}}(d\psi^{(h)})$
has propagator 
$$g^{h}_{\o,\o'}(\xx-\yy)=\int \tilde P_{Z_{h-1}}(d\psi^{(h)})
\psi^-_{\o,\xx}\psi^+_{\o',\yy}$$ 
given by
\begin{eqnarray}
&&{1\over Z_{h-1}}\int d\kk' e^{-i\kk'(\xx-\yy)}{1\over A_{h-1}(k')}\tilde
f_h(\kk')\cdot\nn\\
&&\left(\begin{array}{c c}
-ik_0-\a(k')-v_0\sin k'&
\sigma_{h-1}(k')\\
\sigma_{h-1}(k')&
-ik_0-\a(k')+v_0\sin k'
\end{array}\right)\label{pr11}
\end{eqnarray}
where 
$$ A_{h}(\kk')=
[ -ik_0-\a(k') ]^2 - (v_0\sin k')^2
- [\sigma_{h-1}(\kk')]^2$$
It is convenient to write decompose the propagator as
\begin{equation}\label{klo1}
g^{(h)}_{\o,\o}(\xx-\yy)=
g^{(h)}_{L,\o}(\xx-\yy)+C_2^{(h)}(\xx-\yy)
\end{equation}
where
\begin{equation}\label{klo2}
g^{(h)}_{L,\o}(\xx-\yy)={1\over L\b}\sum_{\kk'} 
{e^{-i\kk'(\xx-\yy)}\over -i k_0+\o v_0\sin k'+\a(k')}\tilde f_h(\kk')
\end{equation}
and, for any integer $N>1$ 
\begin{eqnarray}\label{klo3}
&&|g^{(h)}_{L,\e\o}(\xx-\yy)|\le 
{\g^h C_N\over 1+(\g^h|\xx-\yy|)^N}\\
&&|C_2^{(h)}(\xx-\yy)|\le |{\sigma_h\over \g^h}|^2
{\g^h C_N\over 1+(\g^h|\xx-\yy|)^N}\nn
\end{eqnarray}
Moreover
\begin{equation}\label{klo4}
|g^{(h)}_{\o,-\o}(\xx-\yy)|\le |{\sigma_h\over \g^h}|
{\g^h C_N\over 1+(\g^h |\xx-\yy)|)^N}
\end{equation}

Finally we {\it rescale} the fields so that
\begin{equation}
\int P_{Z_{h-1}}(d\psi^{(\le h-1)}) \int \tilde
P_{Z_{h-1}}(d\psi^{(h)})
\, e^{-\hat\VV^{(h)}
(\sqrt{Z_{h-1}}\psi^{(\le h)})}
\end{equation}
so that
\begin{equation}\label{2.27}
\LL\hat\VV^{(h)}(\psi)=
\g^h\nu_h F_\nu^{(\le h)}+\d_h
F_\a^{(\le h)}+U_h F_U^{(\le h)}
\end{equation}
where by definition
\begin{equation}
\nu_h={Z_h\over Z_{h-1}}n_h;\quad
\d_h={Z_h\over Z_{h-1}}(a_h-z_h);\quad U_h=({Z_h\over Z_{h-1}})^2 u_h
\end{equation}

We perform the integration
\begin{equation}\label{ml}
\int \tilde P_{Z_{h-1}}(d\psi^{(h)}) \, e^{-\hat\VV^{(h)}
(\sqrt{Z_{h-1}}\psi^{(\le h)})}
= e^{-\VV^{(h-1)}(\sqrt{Z_{h-1}}\psi^{(\le h-1)})}\nn
\end{equation}
where $\VV^{(h-1)}$ has the same form as $\VV^{(h)}$ and the procedure
can be iterated, as inserting (\ref{ml}) in (\ref{ml1})
we have an expression like (\ref{3.22})
with $h-1$ replacing $h$. The above procedure is iterated untill
a scale $h^*$ defined as the minimum $h$ such that $\g^{h}>|\s_h|$.
Then we will integrate directly the field
$\psi^{(<h^*)}=\sum_{k=-\io}^{h^*}
\psi^{(k)}$ without splitting the corresponding
integration in scales (as was done
for $h>h^*$). This can be done as $g^{\le h^*}(\xx-\yy)$
verifies the bound eq(\ref{klo3}) with $h^*$ replacing $h$ \ie
it verifies the bound valid for a single scale; 
the reason is that
for momenta larger than $O(\g^{h^*})$ the theory is essentially a massless
theory and for momenta smaller is a massive theory with mass
$O(\g^{h^*})$.
We will call {\it running coupling
constants} $\vec v_h=(U_h,\d_h,\nu_h)$ 
and {\it renormalization constants} $Z_h,\s_h$; their behaviour
as a function of $h$ can be found by an iterative equation
called {\it beta function} (see below).

\subsection{Bounds for the effective potential}
\vskip.5cm
As explained, for example, in [BM], one can write the effective potential
on
scale $j$, if $h\le j<0$, as a sum of terms, which is indeed a finite sum
for
finite values of $N,L,\b$. Each term of this expansion is associated with
a
{\it tree} in the following way.
\vskip5cm

1) Let us consider the family of all trees which can be constructed by
joining a point $r$, the {\it root}, with an ordered set of $b\ge 1$
points,
the {\it endpoints} of the {\it unlabeled tree}, so
that $r$ is not a branching point. $n$ will be called the {\it order} of
the
unlabeled tree and the branching points will be called the {\it non
trivial
vertices}. The unlabeled trees are partially ordered from the root to the
endpoints in the natural way; we shall use the symbol $<$ to denote the
partial order.
 
Two unlabeled trees are identified if they can be superposed by a suitable
continuous deformation, so that the endpoints with the same index
coincide.
It is then easy to see that the number of unlabeled trees with $b$
end-points
is bounded by $4^b$.
 
We shall consider also the {\it labeled trees} (to be called simply trees
in
the following); they are defined by associating some labels with the
unlabeled
trees, as explained in the following items.
 
2) We associate a label $j\le 0$ with the root and we denote $\TT_{j,b}$
the corresponding set of labeled trees with $n$ endpoints. Moreover, we
introduce a family of vertical lines, labeled by an an integer taking
values
in $[j,2]$, and we represent any tree $\t\in\TT_{j,n}$ so that, if $v$ is
an
endpoint or a non trivial vertex, it is contained in a vertical line with
index $h_v>h$, to be called the {\it scale} of $v$, while the root is on
the
line with index $j$. There is the constraint that, if $v$ is an endpoint,
$h_v>j+1$.
 
The tree will intersect in general the vertical lines in set of
points different from the root, the endpoints and the non trivial
vertices;
these points will be called {\it trivial vertices}. The set of the {\it
vertices} of $\t$ will be the union of the endpoints, the trivial vertices
and the non trivial vertices.
Note that, if $v_1$ and $v_2$ are two vertices and $v_1<v_2$, then
$h_{v_1}<h_{v_2}$.
 
Moreover, there is only one vertex immediately following the root, which
will
be denoted $v_0$ and can not be an endpoint; its scale is $j+1$.

3) With each endpoint $v$ of scale $h_v\not=2$ we associate one of the two local
terms contributing to $\LL\hat\VV^{(h_v-1)}$ 
in the r.h.s.
of
(\ref{2.27}) and one space-time point $\xx_v$. 
If $h_v=2$ we associate one of the addend of $\VV$ (\ref{la1}).
We denote by $V_a$ the set of end-points $v$ with $h_v=2$
to which is associated $e^{2i n_v p_F x}\hat\f_{n_v}$ \ie
one of the addenda of $P$ in (\ref{la1}); moreover $q=|V_a|$.

Moreover, we impose the constraint that, if $v$ is an endpoint,
$h_v=h_{v'}+1$, if $v'$ is the non trivial vertex immediately preceding
$v$ and $h_v\not=1$. Given a vertex $v$, $\xx_v$ denotes
the family of all space-time points associated with one end-points
following $v$.  
 
4) If $v$ is not an endpoint, the {\it cluster } $L_v$ with frequency
$h_v$
is the set of endpoints following the vertex $v$; if $v$ is an endpoint,
it is
itself a ({\it trivial}) cluster. The tree provides an organization of
endpoints into a hierarchy of clusters.
 
5) We introduce a {\it field label} $f$ to distinguish the field variables
appearing in the terms associated with the endpoints as in item 3); the
set of
field labels associated with the endpoint $v$ will be called $I_v$.
Analogously, if $v$ is not an endpoint, we shall call $I_v$ the set of
field
labels associated with the endpoints following the vertex $v$; $\xx(f)$,
$\s(f)$ and $\o(f)$ will denote the space-time point, the $\s$ index and
the
$\o$ index, respectively, of the field variable with label $f$.

6) We associate with any vertex $v$ 
a set $P_v$ with the labels associated with the {\it external fields}
of $v$.
These sets must satisfy various
constraints. First of all, if $v$ is not an endpoint and
$v_1,\ldots,v_{s_v}$
are the $s_v$ vertices immediately following it, then $P_v \subset \cup_i
P_{v_i}$
. We shall denote $Q_{v_i}$ the
intersection of $P_v$ and $P_{v_i}$; this definition implies that
$P_v=\cup_i
Q_{v_i}$. The subsets $P_{v_i}/Q_{v_i}$, whose union will be made, by
definition, of the {\it internal fields} of $v$, have to be non empty, if
$s_v>1$, that is if $v$ is a non trivial vertex.
To each $v$ is associated an integer $N_v$
such that
$$\sum_{f\in P_v}\e(f)\kk(f)=2 N_v p_F$$

Given $\t\in\TT_{j,n}$, there are many possible choices of the subsets
$P_v$,
$v\in\t$, compatible with the previous constraints; let us call $\bP$ one
of
this choices. Given $\bP$, we consider the family $\cal G_\bP$ of all
connected Feynman graphs, such that, for any $v\in\t$, the internal fields
of
$v$ are paired by propagators of scale $h_v$, so that the following
condition
is satisfied: for any $v\in\t$, the subgraph built by the propagators
associated with all vertices $v'\ge v$ is connected. The sets $P_v$ have,
in
this picture, the role of the external legs of the subgraph associated
with
$v$. The graphs belonging to $\cal G_\bP$ will be called {\it compatible
with
$\bP$} and we shall denote $\PP_\t$ the family of all choices of $\bP$
such
that $\cal G_\bP$ is not empty.
 
\*
As explained in detail in [BM], we can write, if $h\le j\le -1$,
\begin{eqnarray}\label{2.33}
&&\VV^{(j)}(\sqrt{Z_j}\psi^{[h,j]}) + L\b \tilde E_{j+1}=\nn\\
&&= \sum_{b=1}^\io\sum_{\t\in\TT_{j,b}} \sum_{\bP\in\PP_\t}
\sqrt{Z_j}^{|P_{v_0}|}\int d\xx_{v_0} \tilde\psi^{[h,j]}
(P_{v_0}) K_{\t,\bP}^{(j+1)}(\xx_{v_0})
\end{eqnarray}
where
$$\tilde\psi^{[h,j]}(P_v)= \prod_{f\in P_v}
\psi^{[h,j]\s(f)}_{\xx(f),\o(f)}$$
and $K_{\t,\bP}^{(j+1)}(\xx_{v_0})$ is a suitable function, which is
obtained
by summing the values of all the Feynman graphs compatible with $\bP$, see
item 6) above, and applying iteratively in the vertices of the tree,
different
from the endpoints and $v_0$, the $\RR$-operation, starting from the
vertices
with higher scale; see again [BM] for a more precise definition. 

In order to control, uniformly in $L$ and $\b$, the various sums in
(\ref{2.33}), one has to exploit in a careful way the $\RR$ operation acting
on
the vertices of the tree, as explained in full detail in [BM]. 
The result of
this analysis, which can be applied without any problem to the model
studied
in this paper, is a general bound which has a simple dimensional
interpretation 
and can be easily generalized to bound the 
density
Fourier transforms, as we shall prove in the following sections.
 
Before giving a brief description of this representation, let us see what
happens if we erase the $\RR$ operation in all the vertices of the tree.
In
this case one gets the bound
\begin{eqnarray}
&&\int d\xx_{v_0} |K_{\t,\bP}^{(j+1)}(\xx_{v_0})| \le\\
&&L\b\ (C\bar\e)^{b-q}|\s|^{q} \g^{-j(-2+|P_{v_0}|/2)}\prod_{\rm v\ not\ e.p}
({Z_{h_v}\over Z_{h_v-1}})^{|P_v|\over 2} \g^{-(-2+{|P_v|\over 2}+
{\tilde z(P_v)\over 2})}\prod_{v\in V_a}{|\hat\f_{n_v}|\over|\s|}\nn 
\label{2.36}
\end{eqnarray}
where $C$ is a suitable constant and $\bar\e=
\max_{j+1\le j'\le 0}|\vec
v_{j'}|$; moreover 
$\tilde z(P_v)=1$ if $|P_v|=2$, $\o(f_1)=-\o(f_2)$
and
$\sum_{f\in P_v}\e(f)\o(f)+2 N_v p_F=0$. The factor $\g^{\tilde z(P_v)}$ derives
from the following inequality (see 3.106 of [BM]) if $k>h$
\begin{equation}\label{106}
{|\s_k|\over\g^k}\le {|\s_h|\over\g^h}\g^{{1\over 2}(h-k)}\le \g^{{1\over
2}(h-k)}.
\end{equation}

The good dependence on the number of
end-points $b$ derives from 
the use of the {\it Gram-Hadamard} inequality
for the determinants appearing in 
$K_{\t,\bP}^{(j+1)}(\xx_{v_0})$, see [BM] .
The bound (\ref{2.36}) allows to associate a factor $\g^{2-|P_v|/2+{\tilde z(P_v)\over 2}}$
with any
trivial or non trivial vertex of the tree. This would allow to control the
sums over the scale labels and $\PP_\t$, if $|P_v|$ were larger than $4$
in
all vertices, which is not true. The effect of the $\RR$ operation is to
improve the bound, so that there is a factor less than $1$ associate even
with
the vertices where $|P_v|$ is equal to $2$ or $4$
and eq(\ref{f1}) holds.
In order to explain how
this works, we need a more detailed discussion of the $\RR$ operation.  
We can consider for istance the case of $|P_v|=4$; then
\begin{equation}
\LL \int d\xxx W(\xxx) \prod_{i=1}^{4} \psi^{[h,j]\s_i}_{\xx_i,\o_i}=
\int d\xxx W(\xxx) \prod_{i=1}^{4}
\psi^{[h,j]\s_i}_{\xx_4,\o_i}\big]\;,\label{2.37}
\end{equation}
where $\xxx=(\xx_1,\ldots,\xx_4)$ and $W(\xxx)$ is the Fourier transform
of
$\hat W_{4,\oo}^{(j)}(\kk_1,\kk_2,\kk_3)$.
Note that $W(\xxx)$ is translation invariant (despite the theory
is not translationally invariant); hence
$\psi^{[h,j]\s_i}_{\xx_4,\o_i}$ in the r.h.s. of (\ref{2.37}) can be
substituted with $\psi^{[h,j]\s_i}_{\xx_k,\o_i}$, $k=1,2,3$ and we have
four
equivalent representations of the localization operation, which differ by
the
choice of the {\it localization point}.
 
We have
\begin{eqnarray}
&&\RR \int d\xxx W (\xxx) \prod_{i=1}^{4} \psi^{[h,j]\s_i}_{\xx_i,\o_i}=
\int d\xxx W(\xxx) \left[\prod_{i=1}^{4} \psi^{[h,j]\s_i}_{\xx_i,\o_i}-
\prod_{i=1}^{4} \psi^{[h,j]\s_i}_{\xx_4,\o_i}\right]=\nn\\
&&=\int d\xxx W(\xxx) \Big[
\psi^{[h,j]\s_1}_{\xx_1,\o_1} \psi^{[h,j]\s_2}_{\xx_2,\o_2}
D^{1[h,j]\s_3}_{\xx_3,\xx_4,\o_3} \psi^{[h,j]\s_4}_{\xx_4,\o_4}+
\psi^{[h,j]\s_1}_{\xx_1,\o_1}
D^{1[h,j]\s_2}_{\xx_2,\xx_4,\o_2} \psi^{[h,j]\s_3}_{\xx_4,\o_3}
\psi^{[h,j]\s_4}_{\xx_4,\o_4}+\nn\\
&&+D^{1[h,j]\s_1}_{\xx_1,\xx_4,\o_1}\psi^{[h,j]\s_2}_{\xx_4,\o_2}
\psi^{[h,j]\s_3}_{\xx_4,\o_3} \psi^{[h,j]\s_4}_{\xx_4,\o_4}\Big]
\label{2.38}
\end{eqnarray}
where (again if $L=\b=\io$)
$$D_{\yy,\xx,\o}^{1[h,j]\sigma}=\psi^{[h,j]\sigma}_{\yy,\o}
-\psi^{[h,j]\sigma}_{\xx,\o}$$
 
The field $D_{\yy,\xx,\o}^{1[h,j]\sigma}$ is dimensionally equivalent to
the
product of $|\yy-\xx|$ and the derivative of the field, so that the bound
of
its contraction with another field variable on a scale $j'<j$ will produce
a
``gain'' $\g^{-(j-j')}$ with respect to the contraction of
$\psi^{[h,j]\sigma}_{\yy,\o}$. On the other hand, each term in the r.h.s.
of
(\ref{2.38}) differs from the term which $\RR$ acts on mainly because one
$\psi^{[h,j]}$ field is substituted with a $D^{1[h,j]}$ field and some of
the
other $\psi^{[h,j]}$ fields are ``translated'' in the localization point.
All
three terms share the property that the field whose $\xx$ coordinate is
equal
to the localization point is not affected by the action of $\RR$.
Analogous consideration can be done for the case
$2n=2$; note that one gets the factor $\g^{2(j-j')}$
if $\o_1=\o_2$ and $\g^{(j-j')}$
if $\o_1=\o_2$.
One obtains a new
representation of the effective potential, in place of (\ref{2.33}), 
such that
\begin{equation}\label{2.42}
\RR\VV^{(j)}(\sqrt{Z_j}\psi^{[h,j]})= \sum_{b=1}^\io\sum_{\t\in\TT_{j,b}} \sum_{\bP\in\PP_\t}
\sum_{\a\in A_{\t,\bP}}\sqrt{Z_j}^{|P_{v_0}|}\int d\xx_{v_0}
D_\a\tilde\psi^{[h,j]}
(P_{v_0}) K_{\t,\bP,\a}^{(j+1)}(\xx_{v_0})
\end{equation}
where $A_{\t,\bP}$ labels a finite set of different terms, of counting
power
$C^n$, and, for any $\a\in A_{\t,\bP}$, $D_\a$ denotes an operator
dimensionally equivalent to a derivative of order $m_\a$. The important
property of (\ref{2.42}) is that
\begin{eqnarray}
&&\int d\xx_{v_0} |K_{\t,\bP,\a}^{(j+1)}(\xx_{v_0})| \le
L\b\ (C\bar\e)^{b-q}|\s|^{q} \\
&&\g^{-j(-2+|P_{v_0}|/2+m_\a)}\prod_{\rm v\ not\
e.p}
({Z_{h_v}\over Z_{h_v-1}})^{|P_v|/2} \g^{-[-2+|P_v|/2+z(P_v)+{\tilde z(P_v)\over 2}]}
\prod_{v\in V_a}{|\hat\f_{n_v}|\over|\s|}\nn 
\label{2.36a}
\end{eqnarray}
where $m_\a\ge z(P_{v_0})$ and $z(P_v)$
is vanishing unless

1)$z(P_v)=1$ if $|P_v|=4$ and (\ref{f1}) is verified

2)$z(P_v)=2$
if $|P_v|=2$ and $\o_1=\o_2$ 
and $z(P_v)=1$
if $|P_v|=2$ and $\o_1=-\o_2$ and (\ref{f1}) is verified.

We will need in the following the following result (proved in [BGPS], [GS],
[BoM]).
 
{\bf Theorem.} {\it There is a constant $\e_0$, such that, if
$|\l|\le \e_0$, then, uniformly in the infrared cutoff,}
\begin{equation}
\l_j=\l+O(\l^2)\;,\qquad \d_j=O(\l^2)\;,\qquad h\le j\le 
-1
\end{equation}
by which we obtain that $\bar\e\le C' |U|$. In order
to sum over the trees we need that 
$-2+|P_v|/2+z(P_v)+{\tilde z(P_v)\over 2}>0$, which is however 
{\it not} true.
We define $V_b$ the set of $v$ not end-points
such that $|P_v|=2,4$ and 
$-2+|P_v|/2+z(P_v)+{\tilde z(P_v)\over 2}\le 0$.
For any $v\in V_b$ it holds that, by definition
\begin{equation}\label{ssd}
|\sum_{f\in P_v}\e(f)\o(f) p_F+2 N_v p_F||_T\ge {2\pi\over
Q}
\end{equation}
so that for any $v\in V_b$
\begin{equation}\label{12}
\g^{h_{v'}}\ge {\pi\over 2 Q}
\end{equation} 
It holds that for any $v\in V_b$
there is at least an end-point $\bar v\in V_a$
in the cluster $L_v$. This 
is proved by contradiction: assume that in $v\in V_b$
there are no end-points $\bar v\in V_a$;
as the end-points $v$ not belonging to $V_a$
are such that
$\sum_{f\in P_v}\e(f)\kk'_f=0$, where $\kk'$
is the momentum measured from the Fermi surface,
then a vertex $v$ not
containing end-points $\in V_b$
is such that $\sum_{f\in P_v}\e(f)\kk'_f=0$, which means $v\not\in V_a$
as the l.h.s. of (\ref{ssd}) is vanishing. 
We consider then a maximal $v\in V_b$,
\ie $v$ such that there is no $\bar v\in V_a$ in $L_v$; 
there is a path ${\cal C}\in \t$ 
(called maximal path)
connecting $v$ with an end
point $\bar v\in V_a$; then by (\ref{12}) 
\begin{equation}\label{j1}
1=Q^{2}Q^{-2}\le Q^2 \prod_{v\in {\cal C}}\g^{-\hat z(P_v)}
\end{equation}
where $\hat z(P_v)=2$ if $v\in {\cal C}$ and zero otherwise.
We consider now the next-to-maximal $v\in V_b$ \ie
not belonging to maximal paths and
such that there  
is only one $\bar v\in V_b$ such that $\bar v<v$;
again 
there is a next to maximal path ${\cal C}\in \t$ connecting $v$ with an end
point $\bar v\in V_a$ (not belonging to the maximal paths)
and we obtain again a bound like
(\ref{j1}). We proceed in this way
untill all the $v\in V_b$ are on paths and at the end we get
$$1\le Q^{2q_\t} \prod_{v\in V_b}\g^{-2}$$
so that
\begin{eqnarray}
&&\int d\xx_{v_0} |K_{\t,\bP,\a}^{(j+1)}(\xx_{v_0})| \le
L\b\ (C\bar\e)^{b-q}(Q^2|\s|)^{q} \\
&&\g^{-j(-2+|P_{v_0}|/2+m_\a)}\prod_{\rm v\ not\
e.p}
({Z_{h_v}\over Z_{h_v-1}})^{|P_v|/2} \g^{-[-2+|P_v|/2+
z(P_v)+{\tilde z(P_v)\over 2}+\hat z(P_v)]}
\prod_{v\in V_a}{|\hat\f_{n_v}|\over|\s|}\nn 
\label{2.36b}
\end{eqnarray}
where $\hat z(P_v)=2$ if $v\in V_b$.
Proceding then as in [BM] the sum over $\t$ is convergent
provided that 
\begin{equation}
|\sqrt{\s} Q^2|\le\e\label{a1}
\end{equation}
\vskip.5cm
{\bf Remark:} The above argument does not interfere in the determinant 
estimates, which has to be performed in 
the coordinate space. In fact as all the terms
in the truncated expectations
not verifying (\ref{12}) are exactly vanishing
we can "freely" insert in the sums (before
performing the bounds) a $\chi$-function
ensuring (\ref{12}) (see [BM]).

\subsection{Bounds for the density}

We start from the generating function
and we perform iteratively the integration of the $\psi$ variables;
after the fields $\psi^{(0)},...\psi^{(j+1)}$ have been integrated, we
can write
$$e^{{\cal W}(\phi,J)}=e^{-L\b E_j}
\int P_{\tilde Z_j,C^\e_{h,j}} (d\psi^{[h,j]})
e^{-\VV^{(j)}(\sqrt{Z_j}\psi^{[h,j]})+\BB^{(j)}
(\sqrt{Z_j}\psi^{[h,j]},\phi,J)}$$
where $\BB^{(j)}(\sqrt{Z_j}\psi, \phi,J)$ denotes the sum over the terms
containing at least one $\phi$ or $J$ field; we shall write it in the form
$$\BB^{(j)}(\sqrt{Z_j}\psi, \phi) = \BB_\phi^{(j)}(\sqrt{Z_j}\psi)
+ W_R^{(j)}(\sqrt{Z_j}\psi,\phi,J)$$
where $\BB_\phi^{(j)}(\psi)$ denote the sums over
the terms containing only one $\phi$.

In order to control the Schwinger functions expansion, we have to suitably
regularize $\BB_\phi^{(j)}(\sqrt{Z_j}\psi)$. 
We want to show that, if $j\le -1$, it
can be written in the form
\begin{eqnarray}\label{36}
&&\BB_\phi^{(j)}(\sqrt{Z_j}\psi) =
\sum_\o \sum_{i=j+1}^0 \int d\xx d\yy\;\cdot\nn\\
&&\left[
\phi^+_{\xx,\o} g^{Q,(i)}_{\o}(\xx-\yy){\dpr\over \dpr\psi^+_{\yy\o}}
\VV^{(j)}(\sqrt{Z_j}\psi) + {\dpr\over \dpr\psi^-_{\yy,\o}}
\VV^{(j)}(\sqrt{Z_{j}}\psi) g^{Q,(i)}_{\o}(\yy-\xx)\phi^-_{\xx,\o} \right]+
\nn\\
&&+ \sum_\o\int {d\kk\over (2\p)^2} \left[ \hat\psi^{[h,j]+}_{\kk,\o} 
\hat Q^{(j+1)}_{\o}(\kk)
\hat\phi_{\kk,\o}^- +\hat\phi^+_{\kk,\o} \hat Q^{(j+1)}_{\o}(\kk)
\hat\psi^{[h,j]-}_{\kk,\o} \right]
\end{eqnarray}
where
$$ \hat g^{Q,(i)}_{\o}(\kk)= \hat g^{(i)}_{\o}(\kk) \hat Q^{(i)}_{\o}(\kk)$$
and $Q^{(j)}_{\o}(\kk)$ is defined inductively by the relations
$$\hat Q^{(j)}_{\o}(\kk)=\hat Q^{(j+1)}_{\o}(\kk) - z_j Z_j
D_\o(\kk) \sum_{i=j+1}^0 \hat g^{Q,(i)}_{\o}(\kk)\;,
\quad \hat Q^{(0)}_{\o}(\kk)=1$$

The second line in (\ref{36}) has a simple interpretation in terms of
Feynman graphs; it is obtained by taking all the graphs contributing to
$\VV^{(j)}(\sqrt{Z_{h}}\psi)$ and, given a single graph, by adding a new
space-time-point $\xx$ associated with a term $\phi_\xx \psi_\xx$ and
contracting the correspondent $\psi$ field with one of the external fields
of the graph through a propagator $\sum_{i=j+1}^0
g^{Q,(i)}_{\o}(\xx-\yy)$. 
Hence, it is very easy to see that (\ref{36}) is
satisfied for $j=-1$. The fact that it is valid for any $j$ follows from
our choice to regularize $\BB_\phi^{(j)}(\sqrt{Z_j}\psi)$ by regularizing
the effective potential in the r.h.s. of 
(\ref{36}), that
is by decomposing $\VV^{(j)}$ as 
$\LL \VV^{(j)}+\RR \VV^{(j)}$. This implies, in particular, that
we have to extract from the effective potential the term proportional to
$z_j$; the corresponding contribution to
$\BB_\phi^{(j)}(\sqrt{Z_j}\psi)$ is then absorbed in the term in the third
line of (\ref{36}), before rescaling 
the field $\psi$ and
performing the integration of the scale $j$ field. It is then easy to check
that (\ref{36}) is satisfied for $j=\bar j+1$, if it is satisfied for
$j=\bar j$ (for more details, see [BM]).
 
Note that, if $\hat g^{(j)}_\o(\kk)\not=0$
$$\hat Q_\o^{(j)}(\kk) = 1 - z_j f^\e_{j+1}(\kk) {Z_j\over
\tilde Z_j(\kk) }\;$$
Hence, the propagator $\hat g^{Q,(i)}_{\o}(\kk)$ is equivalent to
$\hat g^{(i)}_{\o}(\kk)$, as concerns the dimensional bounds.

We can expand the functional 
in terms of trees, as we did for the
effective potential, by suitably modifying the definitions 
adding
a new type of endpoints, to be
called of type $\phi$ and to which associated
the terms in the last line of (\ref{36}).
Moreover we change
the definition of the sets $P_v$ so 
that the set $P_v$ includes 
both the
field variables of type $\psi$ which are not yet contracted in the vertex
$v$, to be called {\it normal external fields}, and those which belong to
a normal  endpoint 
and are contracted with a field variable
belonging to an endpoint of type $\phi$ through a propagator $g^{Q,(h_v)}$,
to be called {\it special external fields} of $v$.

From the expansion of the two point Schwinger function
we obtain an expansion for the density. 
We call $h_1$ and $h_2$, with $h_1\le h_2$ (say) the scales of 
the external propagators $g^{Q,(h_v)}$;
we call moreover 
$\bar v$ the first vertex such that  
both the two end-points of type $\phi$ are connected; of course $\g^{h_{\bar v}}\le \g^{h_1}$. 
With respect
to the expansion for the effective potential,
$\RR=1$ for $v\le\bar v$ with $|P_v|=4$.
We write
\begin{equation}\label{cloe}
\r_{n}=\int d\kk \r_{n}(\kk)=\sum_{h_1,h_2}
\int d\kk f_{h_1}(\kk)f_{h_2}(\kk+2n p_F) \r_n(\kk)
\end{equation}
$\r_n$ is given by a sum over trees; in 
order to sum over the scales of the vertices of the tree
we sum over the difference $h_v-h_{v'}$ between consecutive
vertices and over the scale of a vertex; instead of choosing
$v_0$ like in the preceding section we fix
$\bar v$. 
Dimensionally the bound for (\ref{cloe})
is very similar to the bound for the effective
potential with two external line; the $\phi$ end-points
in (\ref{cloe}) are like two $\n$-vertices, and the integration
over $\kk$ allows to associate to the $\phi$ end points
the dimensional factor $\g^{h_1}\g^{h_2}$.
By can bound then (\ref{cloe}) by an expression similar to
(46), with a total dimensional factor $\g^j$
(without $L\b$ factor). However
$\RR=1$ between
$\bar v$ and $v_0$ so that we cannot sum over the scale assignments.
But we can write $\g^j=\g^j
\g^{j-h_{\bar v}}\g^{-(j-h_{\bar v})}$
and at the end
\begin{eqnarray}\label{cc1a}
&&|\r_{n}|\le\sum_{b=1}^\io\sum_{q\le b}\sum_{h_{\bar v}}
[\g^{h_{\bar v}}\\
&&\sum_{\t\in\TT_{h_{\bar v},b,q}} (C\bar\e)^{b-q}(Q^2|\s|)^{q}
\prod_{\rm v\ not\
e.p}
({Z_{h_v}\over Z_{h_v-1}})^{|P_v|/2} \g^{-D_v}
\sum_{\{n_v\}}\prod_{v\in V_a}{|\hat\f_{n_v}|\over|\s|}\nn 
\end{eqnarray}
where
$$D_v=
-[-2+|P_v|/2+
z(P_v)+{\tilde z(P_v)\over 2}+\hat z(P_v)+z'(P_v)]$$
and $z'(P_v)={1\over 3}$ for $\bar v\le v\le v_0$ and zero otherwise. 
As $D_v>0$ 
the second line of (\ref{cc1a}) is surely bounded by a constant.
There are at most $4 b$
non diagonal propagators so that
$$|n|\le 4 b+1+\sum_{v\in V_a} |n_v|$$
so that there exists a $v^*\in V_a$ such that
$|n_{v^*}|\ge C {|n|\over b}$
so that the sum over $\{ n_v\}$ gives
\begin{eqnarray}\label{cc2}
&&|\r_{n}|\le\sum_{b=1}^\io\sum_{q\le b}\sum_{h_{\bar v}}
\g^{h_{\bar v}}\\
&&({b\over |n|})^N\sum_{\t\in\TT_{h_{\bar v},b,q}} 
(C\bar\e)^{b-q}(Q^2|\s|)^{q}
\prod_{\rm v\ not\
e.p}({Z_{h_v}\over Z_{h_v-1}})^{|P_v|/2} \g^{-D_v}\nn 
\end{eqnarray}

\subsection{The case $|n|>1$}
\subsubsection{The case $q\ge 2$.} By (\ref{cc2})
we get
\begin{equation}\label{cc3}
\sum_{b=2}^\io\sum_{q=2}^\io \e^{b-q}(Q^2|\s|^{1\over 2})^{q}
|\s|^{q\over 2}({\k\over|n|})^N
\le|\s|\sum_{b=2}^\io \e^b ({b\over|n|})^N\le {C_N\e^2|\s|\over |n|^N}
\end{equation}

\subsubsection{The case $q=1$ with at least a non diagonal propagator}

If $q=1$ and there is at least a non diagonal propagator
(such terms are contributing to $\tilde\r_n$) we can improve
the bound (\ref{cc2}) with a factor 
${\s_h\over\g^h}$, by (\ref{klo1}). 
Then we get for $b\ge 2$ (so that there is at least a vertex
$\l,\d,\nu$ and using Theorem 1)
$$\sum_{h_{\bar v}} \g^{h_{\bar v}}{\s_{h_{\bar v}}\over\g^{h_{\bar v}}}
\sum_{b=1}^\io |U|^{b-1}(Q^2|\s|^{1\over 2})^{1}
|\s|^{1\over 2}({b\over|n|})^N$$
$$\le |U|\sum_{h_{\bar v}} \s_{h_{\bar v}}
\sqrt{|\s|}\sum_{b=0}^\io \e^b ({b\over|n|})^N\le {C_N |\s|\over |n|^N}$$
using that (which will be proved in \S 3 below)
$$\sum_{k=0}^{h^*}|\s_k|\le |{1\over \h}|\s|(|\s|^{-\h}-1)|$$

We can write explicitly the contribution with $\k=q=1$
\begin{eqnarray}\label{bo1}
&&\hat\r^{1,1}_n=\sum_{h,h'=h^*}^1 {1\over Z_h Z_{h'}}
\sum_{m\not=0,n}\sum_{\o_1,\o'_1\atop\o_2,\o'_2}
\d_{2 m-\o_1+\o'_1-\o_2+\o'_2,2n}\\
&&\l\hat\f_m
\int d\kk' g^{(h)}_{\o_1,\o'_1}(\kk')
g^{(h')}_{\o_2,\o'_2}(\kk'+(2m +\o'_1-\o_2)p_F)\nn
\end{eqnarray}
where $g^{(1)}(\kk)\equiv g^{(1)}_{\o,\o}(\kk')$ if $\kk=\kk'+\o p_F$.
We find the following bound
\begin{eqnarray}\label{bo2}
&&\sum_{h,h'=h^*}^1{1\over Z_h Z_{h'}}|\int d\kk'
g^{(h)}_{\o_1,-\o_1}(\kk')
g^{(h')}_{\o_2,\o'_2}(\kk'+(2m +\o'_1-\o_2)p_F)|\le \nn\\
&&C
\sum_{h=h^*}^1 [|\s_h\g^{-h}|^2+|\s_h\g^{-h}|]\le C'
\end{eqnarray}
This completes the prof of the first of (\ref{ma}).

\subsubsection{The case $q=1$ with no non diagonal propagator}
Such terms are contributing to $c_{n}$.
We can derive such terms 
by the following functional integral
\begin{equation}
{\partial\over\partial J}
\int P(d\psi) e^{\VV+(\psi,\phi)+\int J \l \hat\f_n e^{i n
x}\psi^+\psi^-}|_{J=0}
\end{equation}
with $|n|\not=1$.
By performing the functional integration it appears that
in the effective potential there are terms of the form
$$J\int d\xx d\xx_1 d\xx_2 w(\xx-\xx_1,\xx-\xx_2) 
\l \hat\f_n e^{i 2np_F x}\psi^+_{\xx_1}\psi^+_{\xx_2}$$
which are dimensionally marginal (the $J$-vertex is dimensionally as a couple of
external
$\psi$-fields). In each tree there is a path ${\cal C}$ connecting
the end-point associated to $J$ with the root;
each $v\in {\cal C}$ can have
dimension greater or {\it equal} to
zero and we call $h_{\bar v}$ is the maximal cluster
with two external lines, whose external momenta
verify 
\begin{equation}
|\sum_{i=1}^2\e_i\kk'_i|=|\sum_i\e_i p_F+2 n p_F|
\end{equation}
The r.h.s. of the above equation cannot be vanishing
and using that $\g^{-h_{\bar v}}\le Q$ then
$$\sum_{h_1\ge h_2\ge...h^*}\le {(\log Q)^q\over q!}$$
and
$$\sum_{q=1}^\io U^{q-1} {(\log Q)^q\over q!}\le \log Q Q^{C U}$$

The first
order contribution is given by
\begin{eqnarray}
&&c_n^{(1)}(\s) =\int d\kk
\Big\{
\tilde g_{1,1}^{(1)}(\kk')\,\tilde
g_{1,1}^{(1)}(\kk'+2n\pp_F)\label{bw}\\
&&+ \sum_{\o=\pm 1} \Big[
\tilde g_{1,1}^{(1)}(\kk')\,\sum_{h=h^*}^0
{1\over Z_h}\tilde g_{\o,\o}^{(h)}(\kk'+2n\pp_F+(1-\o)\pp_F)\nn\\
&&+  \sum_{h=h^*}^0
{1\over Z_h}\tilde g_{\o,\o}^{(h)}(\kk')\,
\tilde g_{1,1}^{(1)}(\kk'+2n\pp_F-(1-\o)\pp_F)\nn\\
&&+ \sum_{h=h^*}^0
{1\over Z_h}\tilde g_{\o,-\o}^{(h)}(\kk')\,\sum_{h'=h^*}^0
{1\over Z_{h'}}\tilde g_{-\o,\o}^{(h')}(\kk'+2n\pp_F)\nn\\
&&+ \sum_{h=h^*}^0
{1\over Z_h}
\tilde g_{\o,\o}^{(h')}(\kk')\,\sum_{h'=h^*}^0
{1\over Z_{h'}}\tilde g_{\o,\o}^{(h')}(\kk'+2n\pp_F) +\nn\\
&&\sum_{h=h^*}^0
{1\over Z_h}\tilde g_{\o,\o}^{(h)}(\kk')\,
\sum_{h'=h^*}^0
{1\over Z_{h'}}\tilde g_{-\o,-\o}^{(h')}(\kk'+(2n+2\o)\pp_F)
\Big]\Big\}\nn
\end{eqnarray}
It is easy to see that for the integrals in the first four
lines of (\ref{bw}) are bounded by a constants, as there is
at least a non diagonal propagator or an ultraviolet one;
for the fifth integral the bound is
\begin{eqnarray}
&&\sum_{h=h^*}^0
|\int d\kk'{1\over Z_h}
\tilde g_{\o,\o}^{(h')}(\kk')\,\sum_{h'=h^*}^0
{1\over Z_{h'}}\tilde g_{\o,\o}^{(h')}(\kk'+2n\pp_F)|\le\nn\\
&&\sum_{h=h^*}^0
|\int d\kk'
\tilde g_{\o,\o}^{(h')}(\kk')\,\sum_{h'=h^*}^0
\tilde g_{\o,\o}^{(h')}(\kk'+2n\pp_F)|\le log Q
\end{eqnarray}
If fact if $h<h'$ and noting that $|h'|\le \log Q$
we find
$$\sum_{h<h'}\g^{2h}\g^{-h}\g^{-h'}=\sum_{|h'|\le\log Q}=\log Q$$
This proves the second of (\ref{ma}).

\subsection{The case $|n|=1$}
\subsubsection{The case $q=0$}
The main difference with 
respect to the case $|n|>1$ is that there are
terms with $q=0$ and containing at least one non diagonal propagator 
(such kind of contribution is of course absent if $|n|>1$ by momenta
conservation). The case $q=\k=0$
gives
\begin{eqnarray}
&&c_1(\s) = \int d\kk 
\Big\{ \sum_{h=h^*}^0{\tilde g_{-1,1}^{(h)}(\kk') \over \s} +
\tilde g_{1,1}^{(1)}(\kk')\,\tilde g_{1,1}^{(1)}(\kk'+2\pp_F)\\
&&+ \sum_{\o=\pm 1} \Big[
\tilde g_{1,1}^{(1)}(\kk')\,\sum_{h=h^*}^0
{1\over Z_h}\tilde g_{\o,\o}^{(h)}(\kk'+(3-\o)\pp_F) +\nn\\
&&
\sum_{h=h^*}^0{1\over Z_h}\tilde g_{\o,\o}^{(h)}(\kk')\,\tilde
g_{1,1}^{(1)}(\kk'+(1+\o)\pp_F)
\Big]\Big\}\nn\\
&&= -F(\s,L,U)+\tilde c_1(\s)\nn
\end{eqnarray}
and more explicitly
\begin{eqnarray}\label{hj}
&&F(\s,\l,U)\equiv \sum_{h=h^*}^0{\tilde g_{-1,1}^{(h)}(\kk') \over \s}=\\
&&\sum_{h=h^*}^0
\int d\kk {1\over Z_h}{\s_h \over \s}{f_h(\kk)\over k_0^2+\sin^2
k'+(1-\cos
k')^2+\s_h^2}\nn
\end{eqnarray}
where $\s_h,Z_h$ verify the flow equation (\ref{rg}) below,
and $|\tilde c_1|\le C$ where $C$ is a constant.
In bounding $q=0$, $\k\ge 1$ and at least a non-diagonal
propagator we use (\ref{cc2}) with the improving that
there is at least a ${\s_{h_{\bar v}}\over\g^{h_{\bar v}}}$
more for the presence of the non diagonal propagator;
we get
\begin{eqnarray}\label{cc3}
&&|\r_{n}|\le\sum_{\k=1}^\io\sum_{q\le k}\sum_{h_{\bar v}}
|\s_{h_{\bar v}}|\\
&&\g^{-h_{\bar v}}\sum_{\t\in\TT_{h_{\bar v},\k,q}} (C\bar\e)^{\k-q}(Q^2|\s|)^{q}
\prod_{\rm v\ not\
e.p}
({Z_{h_v}\over Z_{h_v-1}})^{|P_v|/2} \g^{-D_v}
\sum_{\{n_v\}}\prod_{v\in V_a}{|\hat\f_{n_v}|\over|\s|}\nn\\
&&\le 
C_N|U||{|\s|\over\h}[|\s|^{-\h}-1]|\nn 
\end{eqnarray}

\subsubsection{The case $q\ge 1$}
If $q\ge 2$ we proceed as in section 2.4.1 and we get the bound 
$C_N|\s|$. Again, if $q=1$ and there is at least a non
diagonal propagator we get the bound 
$C_N|\s|$ proceding as in 2.4.2.
Finally if $q=1$ and there are no non diagonal
propagators we proceed as in 2.4.3 but 
the clusters $v$ containing $\l\hat\f_{\pm 1}$ with two or
for external lines are such that $N_v=2$ so that $\g^{-h_{v'}}\le {\pi\over
4}$; we proceed then as in 2.4.3 but instead of $Q$ in the bounds
we have an harmless constant and  again we get the bound
$C_N |\s|$.
\vskip.5cm
Then by (\ref{hj})
can be written as
\begin{equation}
{\l^2\over\h}[
({|\s|\over A})^{-\h}-1][a^{-1}+F(\l,U,\s)]+\s f(\l,U,\s)
=1\label{ffa1}
\end{equation}
with $|F|\le C |U|$, $|f|\le C$
and $\h_1=\b_1 U+\tilde\h_1$, $\h_2=\b_1 U+\tilde\h_2$,
$|\tilde\h_1|,|\tilde\h_2|\le C U^2$, and $C$,
$a,\b_1,A$ are positive constants.

The term $F$ in the above equation comes just from from
the terms with $|V|=1$, $q\ge 2$ and no non diagonal propagators. 
In the case the only vertex belonging to $V(\t)$ is
$\l\hat\f_1$ and it is contracted at scale $k=1$; 
by using 
$${\s_1\over\g^1}\le \g^{{1\over 2}(h-1)}{\s_h\over \g^h}$$
we see that we can renormalize all the clusters between
$1$ and $h$ and containing $\l\hat\f_1$, so that
we have a bound
of the form, for $q\ge 1$
\begin{equation}\label{a5}
|\r_{1}^{q,1}(\t)|\le (C\bar\e)^{q-1}{\s_h\over\g^h}\g^h
\prod_{v not e.p.}
\g^{-(-2+|P_{v}|+z(P_{v})+\bar z(P_v))}
\end{equation} 
which gives the bound for $F$.
\vskip.5cm

\section{The flow of the running coupling constants}

\subsection{The Beta function}

The equations for the running coupling constants
are,
for $h\ge h^*$ 
\begin{eqnarray}\label{ll1}
&&\nu_{h-1}=\g\nu_h+G^{h}_\nu\quad U_{h-1}=U_h+G^{h}_U\nn\\
&&\sigma_{h-1}=\sigma_h
+G_{\sigma}^{h}\quad
\d_{h-1}=\d_h+G^{h}_\d\nn\\
&&{Z_{h-1}\over Z_h}=1+G_{z}^{h}
\end{eqnarray}
It is convenient to split $G_i^{(h)}$,
with $i=\mu,\s,\nu$ as
\begin{eqnarray}
&&G^{h}_i(\mu_h,\nu_h,\s_h;...;
\mu_0,\nu_0,\s_0)=\\
&&G^{1,h}_i(\mu_h,\nu_h;...;
\mu_0,\nu_0)+
G^{2,h}_i(\mu_h,\nu_h,\s_h;...;
\mu_0,\nu_0,\s_0)\nn
\end{eqnarray}
where we have splitted $g^{(h)}_{\o,\o}$ as in eq(\ref{klo1})
and 
$G^{1,h}_i$ contains no non diagonal
propagators and only the part $g^{(h)}_{L;\o,\o}$
of the diagonal propagators; moreover
there are no vertices with $m\not=0$; in 
$G^{2,h}_i$ there are all the remaining contributions.
It is easy to check that for $i=\mu,\nu$, for $max_{k\ge h}|\vec
v_k|\le\e$
\begin{equation}\label{111}
|G^{2,h}_i|\le C[{\s_h\over\g^h}]^2\e^2
\end{equation}
This follows from the bound eq(\ref{klo3}) for $C_2^h$ and from the fact
that $\nu,\mu$ are momentum conserving terms.
For $i=\s$ by symmetry reasons, 
$G^{1,h}_i\equiv 0$ and 
\begin{equation}
|G^{2,h}_\s(\mu_h,\nu_h,\s_h;...;
\mu_0,\nu_0,\s_0)|\le C |U_h \s_h|
\end{equation}
We decompose, if $i=\mu,\nu$ 
\begin{eqnarray}\label{ca13}
&&G^{1,h}_i(\mu_h,\nu_h;...;\mu_0,\nu_0)=\nn\\
&&\bar G^{1,h}_i(\mu_h;...;\mu_0)+
\hat G^{1,h}_i(\mu_h,\nu_h;...;\mu_0,\nu_0)
\end{eqnarray}
where the first term in the r.h.s. of eq(\ref{ca13}) is obtained putting
$\nu_k=0$, $k\ge h$ in the l.h.s.
It is easy to see, from the fact that $g^{(h)}_{L,\o}(\xx;\yy)$
can be divided in a even part plus a correction smaller
than a factor $\g^{h\over 4}$, 
for $max_{k\ge h}|\vec v_k|\le\e$
\begin{equation}\label{78}
|\bar G^{1,h}_\nu(\mu_h;...;\mu_0)\le C\e\g^{h\over 4}
\end{equation}
On the other hand
\begin{equation}
|\bar G^{1,h}_\mu(\mu_h;...;\mu_h)|\le C 
\e^2\g^{{h\over 2}}
\end{equation}
as one can prove using the exact solution
of the Luttinger model\cite{[BGPS],[BM1],[GS]}.
Moreover we have that, for $i=\nu,\mu$
\begin{equation}
|\hat G^{1,h}_i(\mu_h,\nu_h;...;\mu_0,\nu_0)|\le
C\n_h |U_h|^2
\end{equation}
Finally by
a second order computation one obtains
\begin{eqnarray}
&&G^{1,h}_{\s}=\s_h U_h [\b_1+\bar G^{1,h}_\s]\nn\\
&&G^{1,h}_z=U^2_h[\b_2+\bar G^{1,h}_z]
\end{eqnarray}
with $\b_1,\b_2$ non vanishing positive constants
and $|\bar G^{1,h}_\s|\le C |U_h|$,
and
$|\bar G^{1,h}_z|\le C |U_h|$.

By using the above properties we can control the flow
of the running coupling constants. In fact, 
if 
$|\nu_k|\le C \e[{\g^{k\over 4}+{|\s_k|\over\g^k}}]
$ for any $k\ge h^*$ (what will be proved
in the following section) it follows that
there exist positive constants
$c_1,c_2,c_3,c_4,C$
such that, if $\l,u$ are small enough and $h\ge h^*$:
\begin{eqnarray}\label{ed}
&&
|U_{h-1}-U|<C U^{3/2}\nn\\
&&e^{-U\b_1 c_3 h}< {|\sigma_{h-1}|\over
|\sigma_0|}< e^{-U\b_1 c_4 h} \label{rg}\\
&&e^{-\b_2 c_1 U^2 h} < Z_{h-1}< e^{-\b_2 c_2 U^2 h}\nn
\end{eqnarray}

\subsubsection{Determination of the counterterm $\nu$}

We show finally that it is possible to fix $\nu$ to a $\l$-independent
value; more exactly we show that it is possible to choose
$\nu$ as in the $\l=0$ case so that $\nu_k$ is small for any 
$k\ge h^*$. In the $\l=0$ case $\s_k=0$
and there are no contribution to the effective potential
with $m\not=0$; calling 
$\tilde\nu_k,\tilde\mu_k$ the analogous of $\nu_k,\mu_k$ 
we can write
\begin{equation}\label{n1}
\tilde\nu_h=\g^{-h+1}[\nu+\sum_{k=h+1}^1\g^{k-2}G^{1,k}_\nu(\tilde\nu,\tilde\mu)]
\end{equation}
where $G^{1,k}_\nu(\tilde\nu,\tilde\mu)=G^{1,k}_\nu(\tilde\nu_{k},
\tilde\mu_{k};....;\tilde\nu_{0},
\tilde\mu_{0})$.
We choose
\begin{equation}
\label{n2}
\nu=-\sum_{k=-\io}^1\g^{k-2}G^{1,k}_\nu(\tilde\nu,\tilde\mu)
\end{equation}
then
\begin{equation}
\tilde\nu_h=-\g^{-h}\sum_{k=-\io}^h \g^{k-1}
G^{1,k}_\nu(\tilde\nu,\tilde\mu)
\end{equation}
and  
$|\tilde\nu_h|\le C\e \g^{{h\over 4}}$
as by (\ref{78}) 
\begin{equation}
\g^{-h}\sum_{k=-\io}^h \g^{k-1}|G^{1,k}_\nu(\tilde\nu,\tilde\mu)|\le C'
\e \g^{-h}\sum_{k=-\io}^h \g^{k}\g^{{1\over 4} k}
\le C\e\g^{h\over 4}
\end{equation}
For the model with $\l\not=0$, for $h\ge h^*$
\begin{equation}\label{n11}\nn
\nu_h=\g^{-h+1}[\nu+\sum_{k=h+1}^1
\g^{k-2}[G^{2,k}_\nu(\nu,\mu,\s)+G^{1,k}_\nu(\nu,\mu)]]
\end{equation}
and inserting $\nu$ given by eq(\ref{n2})
\begin{eqnarray}\label{n111}
&&\nu_h-\tilde\nu_h=\g^{-h+1}\{\sum_{k=h+1}^1
\g^{k-2}G^{2,k}_\nu(\nu,\mu,\s)\nn\\
&&+\sum_{k=h+1}^1\g^{k-2}[G^{1,k}_\nu(\nu,\mu)-  
G^{1,k}_\nu(\tilde\nu,\tilde\mu)]\}
\end{eqnarray}
We prove that, for $h\ge h^*$ 
\begin{equation}\label{ind}
|\nu_h-\tilde\nu_h|\le \e \bar C
({\s_h\over\g^h})^2\qquad|\tilde\mu_h-\mu_h|\le \e ({\s_h\over\g^h})^2
\end{equation}
The proof is done by induction assuming that it
holds for scales $\ge h+1$ and proving
by (\ref{n111}) that it holds for scale $h$. Looking
at the first sum in (\ref{n111}) and using (\ref{111}),(\ref{ed})
\begin{eqnarray}
&&\g^{-h}\sum_{k=h+1}^1
\g^{k-2}|G^{2,k}_\nu(\nu,\mu)|\le C_1 
\g^{-h}\e^2 \sum_{k=h+1}^1 \g^{k-2} ({\s_k\over\g^k})^2\nn\\
&&\le({\s_h\over\g^h})^2 C_2
\e^2 \sum_{k=h+1}^1 \g^{h-k} ({\s_k\over\s_h})^2\le C_3 \e^2
({\s_h\over\g^h})^2
\end{eqnarray}

Finally we can write
$$G^{1,k}_\nu(\nu,\mu)-  
G^{1,k}_\nu(\tilde\nu,\tilde\mu)=\sum_{\bar k>k} D_{\bar k,k}$$
with
\begin{eqnarray}
&&D_{\bar k,k}=G^{1,k}_\nu(\nu_k,\mu_k;...;\nu_{\bar k},\mu_{\bar k};
\tilde\nu_{\bar k+1},\tilde\mu_{\bar
k+1};...;\tilde\nu_0,\tilde\mu_0)-\nn\\
&&G^{1,k}_\nu(\nu_k,\mu_k;...;\tilde \nu_{\bar k},\tilde\mu_{\bar k};
\tilde\nu_{\bar k+1},\tilde\mu_{\bar k+1};...;\tilde\nu_0,\tilde \mu_0)\nn
\end{eqnarray}
and, by the inductive hypothesis and (\ref{rom})
\begin{equation}
\sum_{\bar k\ge k}|D_{\bar k,k}|\le C_1\bar C \e^2 \sum_{\bar k\ge
k}\g^{{1\over2}(k-\bar k)}
({\s_{\bar k}
\over\g^{\bar k}})^2\le C_2\bar C \e^2 ({\s_{k}
\over\g^{k}})^2
\end{equation}
so that the last sum in (\ref{n111}) is bounded by
$\bar C\e({\s_{h}
\over\g^{h}})^2$
with $\bar C=4C$,
and eq(\ref{ind}) is proved; by that equation
it follows that for $h\ge h^*$ then $|\nu_h|\le C\e$, so that it is
possible
to have that the flow of $\nu_h$ is bounded choosing a $\l$ independent
$\nu$.

\subsubsection{Proof that $\h\equiv\h(U)$}

In a similar way we can prove that the critical index
$\h$ is $\l$ independent; remembering that
$${\s_{h-1}\over\s_h}=1+G_1(\l_h,\d_h)+{\s_h\over\g^h}\bar G_2$$
where to $G_1$ are contributing all the terms with
relevant end-points and no non diagonal propagators. 
We introduce
$${\bar\s_{h-1}\over\bar\s_h}=1+G_1(\l_h,\d_h)$$
such that $\bar\s_{h^*}=\s\g^{\h h^*}$ with $\h$ of course
$\s$ independent. We write
$$\s_h=C_h\bar\s_h$$
By simple substitution it is in fact easy to see that
$${c_{h-1}\over c_h}=1+{\bar\s_h\over\bar\s_{h-1}}
[{\s_h\over\g^h}\bar G_2]$$
so that
$${|c_{h-1}|\over |c_h|}\le e^{C\e^2{\s_h\over\g^h}}$$
and
$${|c_{h-1}|\over |c_0|}\le e^{C\e^2\sum_{k=0}^h
{\s_k\over\g^k}}\le 1+O(\e^2)$$

\section{References}

[AAR] S. Aubry, G. Abramovici, J. Raimbault:
{\it J. Stat. Phys.} {\bf  67}, 675--780 (1992)
\vskip.5cm
[BGM] G.Benfatto,G.Gentile, V.Mastropietro.
{\it J. Stat. Phys.} 92, 1071-1113 (1998)
\vskip.5cm
[BM] G.Benfatto, V.Mastropietro: {\it Rev. Math. Phys.} in press
\vskip.5cm
[BoM] F.Bonetto, V.Mastropietro: {\it Commun. Math. Phys.}, 172,
57-93 (1995)
\vskip.5cm
[BGPS]G. Benfatto, G. Gallavotti, A. Procacci, B. Scoppola:
{\it  Comm. Math. Phys.} {\bf  160}, 93--171 (1994)
\vskip.5cm
[F] R.H. Friend: in Solitons and condensed matter physics, Ed A.R.Bishop
and T. Schneder, Springer, (1978)
\vskip.5cm
[FK] J.K. Freericks, M. Jarrell cond mat/9502098 
\vskip.5cm
[GS] G.Gentile, B. Scoppola
{\it Comm. Math. Phys.} {\bf 154}, 153--179 (1993).
\vskip.5cm
[KL] T. Kennedy, E.H. Lieb:
{\it Phys. Rev. Lett.} {\bf 59}, 1309--1312 (1987).
\vskip.5cm
[LN] E.H. Lieb, B. Nachtergale; {\it Phys. Rev. B},51,8,4777-4791
(1995)
\vskip.5cm
[P]
R.E. Peierls: in Quantum Theory of solids
, O.U.P., (1955)
\vskip.5cm
\end{document}